\documentclass[12pt,preprint]{aastex}

\newcommand{\al}{Alfv\'en}
\begin{document}
\title{The Inability of Ambipolar Diffusion to set a Characteristic
  Mass Scale in Molecular Clouds}
\author{Jeffrey S. Oishi\altaffilmark{1,2}}
\email{joishi@amnh.org}
\author{Mordecai-Mark Mac Low\altaffilmark{2}}
\email{mordecai@amnh.org}

\altaffiltext{1}{Department of Astronomy, University of Virginia,
  P.O. Box 3818, Charlottesville, VA 22903} 
\altaffiltext{2}{Department of Astrophysics, American Museum of
  Natural History, New York, NY 10024-5192} 
\begin{abstract}

  We investigate the question of whether ambipolar diffusion
  (ion-neutral drift) determines the smallest length and mass scale on
  which structure forms in a turbulent molecular cloud. We simulate
  magnetized turbulence in a mostly neutral, uniformly driven,
  turbulent medium, using a three-dimensional, two-fluid,
  magnetohydrodynamics (MHD) code modified from Zeus-MP. We find that
  substantial structure persists below the ambipolar diffusion scale
  because of the propagation of compressive slow MHD waves at smaller
  scales.  Contrary to simple scaling arguments, ambipolar diffusion
  thus does not suppress structure below its characteristic
  dissipation scale as would be expected for a classical diffusive
  process.  We have found this to be true for the magnetic energy,
  velocity, and density.  Correspondingly, ambipolar diffusion leaves
  the clump mass spectrum unchanged.  Ambipolar diffusion appears
  unable to set a characteristic scale for gravitational collapse and
  star formation in turbulent molecular clouds.
\end{abstract}


\section{Introduction}
Molecular clouds are turbulent, with linewidths indicating highly
supersonic motions \citep{1974ARA&A..12..279Z}, and magnetized, with
magnetic energies in or near equipartition with thermal energy
\citep{1999ApJ...520..706C}.  They have low ionization fractions
\citep{e79} leading to imperfect coupling of the magnetic field with
the gas.  Molecular clouds are the sites of all known star formation,
so characterizing the properties of this non-ideal, magnetized
turbulence appears central to formulating a theory of star formation.

The drift of an ionized, magnetized gas through a neutral gas coupled
to it by ion-neutral collisions is known by astronomers as ambipolar
diffusion (AD) and by plasma physicists as ion-neutral drift. It was
first proposed in an astrophysical context by
\citet{1956MNRAS.116..503M} as a mechanism for removing magnetic flux
and hence magnetic pressure from collapsing protostellar cores in the
then-novel magnetic field of the Galaxy. However, more recently, as
turbulence has regained importance in the theory of star formation, AD
has been invoked as a source of dissipation for magnetic energy in the
turbulent magnetohydrodynamic (MHD) cascade and thus a characteristic
length scale for the star formation process
\citep*[e.g.,][]{2004ApJ...616..283T}. 
This is due to its well-known ability to damp certain families of
linear MHD waves
\citep*{1969ApJ...156..445K,1988ApJ...332..984F,1996ApJ...465..775B}.
However, as \citet{1996ApJ...465..775B} pointed out, AD does allow
slow modes to propagate undamped.

A brief calculation suggests that AD should be the most important
dissipation mechanism in molecular clouds. AD can be expressed as an
additional force term in the momentum equation for the ions
\begin{equation} \label{force_in}
F_{in} = \rho_i \rho_n \gamma_{AD} (\mathbf{v_n} - \mathbf{v_i}),
\end{equation} 
and an equal and opposite force $F_{ni} = - F_{in}$ in the neutral
momentum equation, where $\rho_i$ and $\rho_n$ are the ion and neutral
densities and $\gamma_{AD} \simeq 9.2 \times 10^{13}\ \mathrm{cm^3\
s^{-1}\ g^{-1}}$ is the collisional coupling constant \citep*{1983ApJ...264..485D,1997A&A...326..801S}.

The effect of ion-neutral drift on the magnetic field can be simply
expressed in the strong coupling approximation
\citep{1983ApJ...273..202S} that neglects the momentum and pressure of
the ion fluid and equates the collisional drag force on the ions
$F_{in}$ with the Lorentz force,
\begin{equation} \label{strong}
-\rho_i \rho_n \gamma_{AD} (\mathbf{v_i} - \mathbf{v_n}) = \frac{\mathbf{(\nabla \times B) \times B}}{4 \pi}.
\end{equation}
\citet{1994ApJ...427L..91B} note that by substituting
equation~(\ref{strong}) into the induction equation for the ions, one
arrives at
\begin{equation} \label{induction}
\partial_t \mathbf{B} = \mathbf{\nabla} \times \left[(\mathbf{v_n \times
    B})+ \frac{\mathbf{(\nabla \times B) \cdot B}}{4 \pi \rho_i \rho_n
    \gamma_{AD}} \mathbf{B} - (\eta + \eta_{AD}) \mathbf{\nabla \times
    B}\right],
\end{equation}
where 
\begin{equation} 
\eta_{AD} = \frac{B^2}{4 \pi \rho_i \rho_n \gamma_{AD}}
\end{equation} 
is the ambipolar diffusivity and $\eta$ is the Ohmic diffusivity.
However, dissipation is not the only contribution of AD to the
induction equation. Given that AD tends to force magnetic fields into
force-free states \citep{1995ApJ...446..741B,1997ApJ...478..563Z} with
$\mathbf{(\nabla \times B) \times B} = 0$, it should come as little
surprise that the $\mathbf{(\nabla \times B ) \cdot B}$ term must be
given proper consideration.

We can approximate the scale $\ell_{ds}$ below which dissipation
dominates turbulent structure for a given diffusivity $\eta$ in at
least two ways. The first is commonly used in the turbulence
community. It is to equate the driving timescale
\begin{equation} \label{taudr}
\tau_{dr}  = L_{dr} / v_{dr}, 
\end{equation}
where $L_{dr}$ is the driving wavelength and $v_{dr}$ is the rms
velocity at that wavelength, with the dissipation timescale $\tau_{ds}
= \ell_{ds}^2 / \eta$, and solve for $\ell_{ds}$.  The second method
was suggested by \citet{1996ApJ...465..775B} and
\citet{1997ApJ...478..563Z} and advocated by \citet{khm00}. It is to
estimate the length scale at which the Reynolds number associated with
a given dissipation mechanism becomes unity. The Reynolds number for
ion-neutral drift can be defined as
\begin{equation}
R_{AD} = \frac{L V}{\eta_{AD}},
\end{equation}
where $V$ is a characteristic velocity. This method requires setting $R_{AD}$ to one and solving for
$L = \ell_{ds}$ to find
\begin{equation}
\ell_{AD} = \frac{B^2}{4 \pi \rho_i \rho_n \gamma_{AD} V}.
\end{equation}

\citet{khm00} show that by adopting values characteristic of dense
molecular clouds, a magnetic field strength $B= 10 B_{10}\ \mathrm{\mu
G}$, ionization fraction $x = 10^{-6} x_6$, neutral number density
$n_n = 10^3 n_3\ \mathrm{cm^{-3}}$, mean mass per particle $\mu = 2.36
m_H$ where $m_H$ is the hydrogen mass, such that $\rho_n = \mu n_n$,
and the above value for the ion-neutral coupling constant, the length
scale at which AD is important is given by
\begin{equation}
\ell_{AD} = (0.04\ \mathrm{pc}) \frac{B_{10}}{M_A x_6 n_3^{3/2}},
\end{equation}
where $M_A = V/v_A$ is the \al\ Mach number. By contrast, Ohmic
dissipation acts only at far smaller scales, $\ell_\eta \sim 10^{-13}\ 
\mathrm{pc}$ \citep{1997ApJ...478..563Z}.

For our purposes, we use the Reynolds number method and choose $V =
v_{RMS}$, the RMS velocity. Although we use Reynolds numbers, we find
that using the timescale method has no effect on our results.

Previous three-dimensional numerical studies of turbulent ion-neutral
drift have used the strong coupling approximation
\citep*{2000ApJ...540..332P}. This by definition renders simulations
unable to reach below $R_{AD} \sim 1$, and thus into the dissipation
region.

In this paper, we present runs in which we vary the ambipolar
diffusion coupling constant, and thus $\ell_{AD}$.  We find a
surprising lack of dependence of the spectral properties on the
strength of the ambipolar diffusivity.  In particular, no new
dissipation range is introduced into the density, velocity or magnetic
field spectra by ambipolar diffusion, nor is the clump mass spectrum
materially changed.

\section{Numerical Method}

We solve the two-fluid equations of MHD using the ZEUS-MP code
\citep{2000RMxAC...9...66N} modified to include a semi-implicit
treatment of ion-neutral drift. ZEUS-MP is the domain-decomposed,
parallel version of the well-known shared memory code ZEUS-3D
\citep{cn94}. Both codes follow the algorithms of ZEUS-2D
\citep{1992ApJS...80..753S,1992ApJS...80..791S}, including
\citet{1977JCoPh..23..276V} advection, and the constrained transport
method of characteristics
\citep{1988ApJ...332..659E,1995CoPhC..89..127H} for the magnetic
fields. We add an additional neutral fluid and collisional coupling
terms to both momentum equations. Because ion-neutral collisions
constitute a stiff term, we evaluate the momentum equations using the
semi-implicit algorithm of \citet{1997ApJ...491..596M}.  We also
include an explicit treatment of Ohmic diffusion by operator splitting
the induction equation \citep*{2000ApJ...530..464F}.

We ignore ionization and recombination, assuming that such processes
take place on timescales much longer than the ones we are concerned
with. This means that ions and neutrals are separately
conserved. Furthermore, we assume that both fluids are isothermal and
at the same temperature, thus sharing a common sound speed $c_s$.

\subsection{Initial Conditions and Parameters}

All of our runs are on three-dimensional Cartesian grids with periodic
boundary conditions in all directions. 

The turbulence is driven by the method detailed in
\citet{1999ApJ...524..169M}. Briefly, we generate a top hat function
in Fourier space between $1 < |k| < 2$. The amplitudes and phases of
each mode are chosen at random, and once returned to physical space,
the resulting velocities are normalized to produce the desired RMS
velocity, unity in our case. At each timestep, the same pattern of
velocity perturbations is renormalized to drive the box with a
constant energy input ($\dot{E} = 1.0$ for all simulations) and applied
to the neutral gas.

Our isothermal sound speed is $c_s = 0.1$, corresponding to an initial
RMS Mach number $M = 10$. The initial neutral density $\rho_n$ is
everywhere constant and set to unity. The magnetic field strength is
set by requiring that the initial ratio of gas pressure to magnetic
pressure be everywhere $\beta = 8 \pi c_s^2 \rho / B^2 = 0.1$; its
direction lies along the z-axis.

Although our semi-implicit method means that the timestep is not
restricted by the standard Courant condition for diffusive processes
(that is, $\propto [\Delta x]^2$), the two-fluid model is limited by
the \al\ timestep for the ions. This places strong constraints on the
ionization fraction ($x = n_i/n_n$) we can reasonably compute. We
therefore adapt a fixed fraction of $x= 0.1$ for our simulations.
While this fraction is certainly considerably higher than the
$10^{-4}$--$10^{-9}$ typical of molecular clouds, the ionization
fraction only enters the calculation in concert with the collisional
coupling constant $\gamma_{AD}$. Thus, we are able to compensate for
the unrealistically high ionization fraction by adjusting
$\gamma_{AD}$ accordingly.

We present four runs, two with AD, one with Ohmic diffusion, and one
ideal MHD run (see Table~\ref{run_tab}). For the AD runs, we vary the
collisional coupling constant in order to change the diffusivity. 

Our results are reported for a resolution of $256^3$ at time $t =
0.125 t_s = 2.5$ where $t_s = 20$ is the sound crossing time for the
box. This exceeds by at least 30\% the turbulent crossing time over
the driving scale $\tau_{dr}$ computed from equation~(\ref{taudr}),
and tabulated in Table~\ref{run_tab}. Our computation of $\tau_{dr}$
is done for $L_{dr} = 1$, the maximum driving wavelength.
\citet*{2003ApJ...590..858V} note that $\tau_{dr}$ is the relevant
timescale for the formation of nonlinear structures. Furthermore, we
find from studies performed at $128^3$ out to $t = 0.3 t_s$ that
$0.125 t_s$ is enough time to reach a steady state in energy.

\section{Results}

\begin{figure} 
\plotone{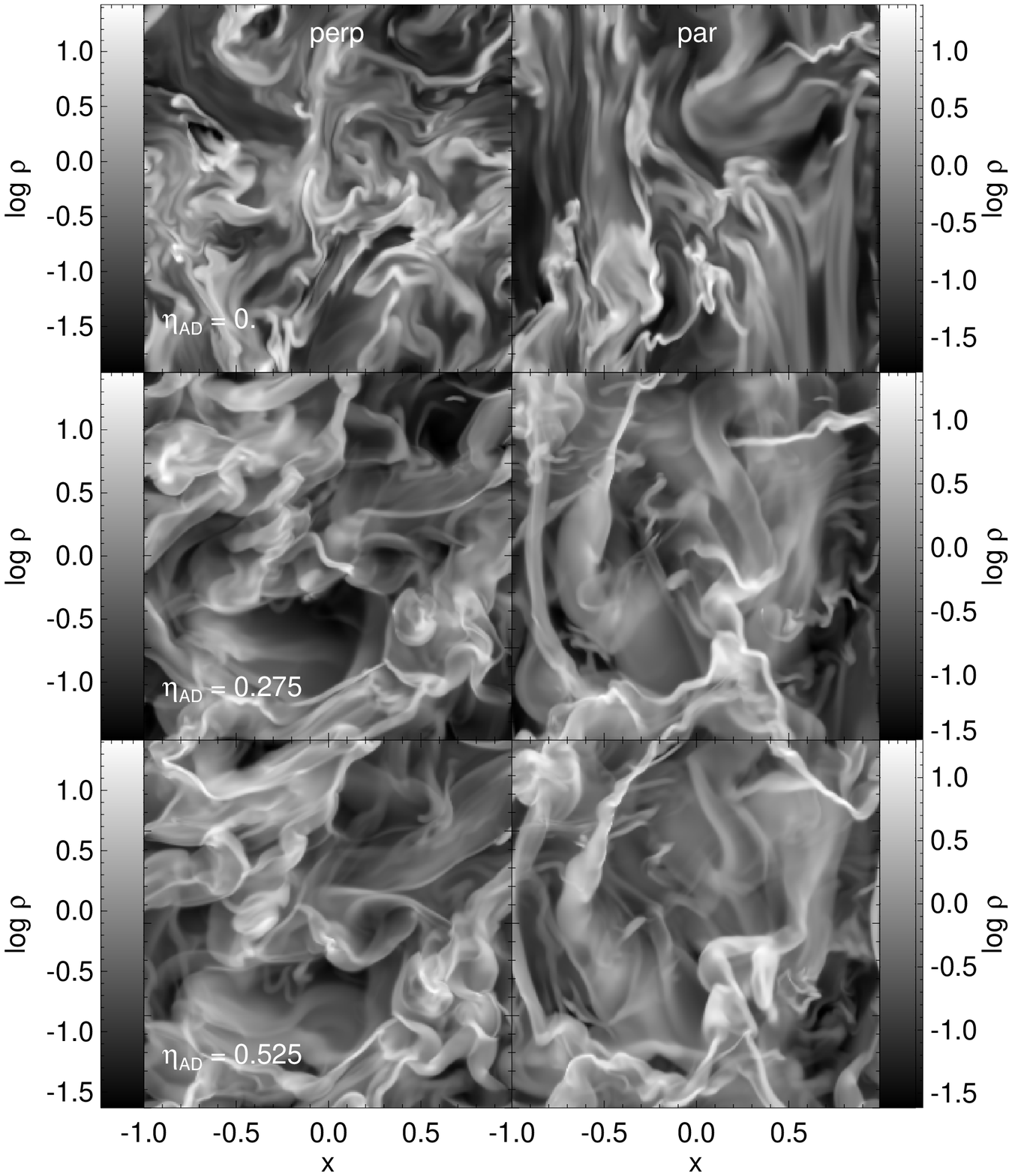}
\caption{Random cuts of density $\rho$ parallel and perpendicular to the
  magnetic field for each of three runs of varying 
  ambipolar diffusivity $\eta_{AD}$. Each  image is scaled to  its own
  minimum and maximum,  enhancing  structural features. For  AD  runs,
  $\rho = \rho_i + \rho_n$ \label{rho_pic}}
\end{figure}

Figure \ref{rho_pic} shows cuts of density perpendicular and parallel
to the mean magnetic field. For the ambipolar diffusion runs, we show
the total density $\rho = \rho_i + \rho_n$. The morphology of density
enhancements in the different runs appears similar, giving a
qualitative suggestion of the quantitative results on clump mass
spectra discussed next.

\subsection{Clump mass spectrum}
We wish to understand whether AD determines the smallest scale at
which clumps can form in turbulent molecular clouds. Determining
structure within molecular clouds has proved difficult in both theory
and observation. Molecular line maps (eg, Falgarone et al 1992) show
that for all resolvable scales, the density fields of clouds is made
up of a hierarchy of clumps. Furthermore, the identification of clumps
projected on the sky with physical volumetric objects is questionable
\citep*{2001ApJ...546..980O,2002ApJ...570..734B}.

\begin{figure} 
\plotone{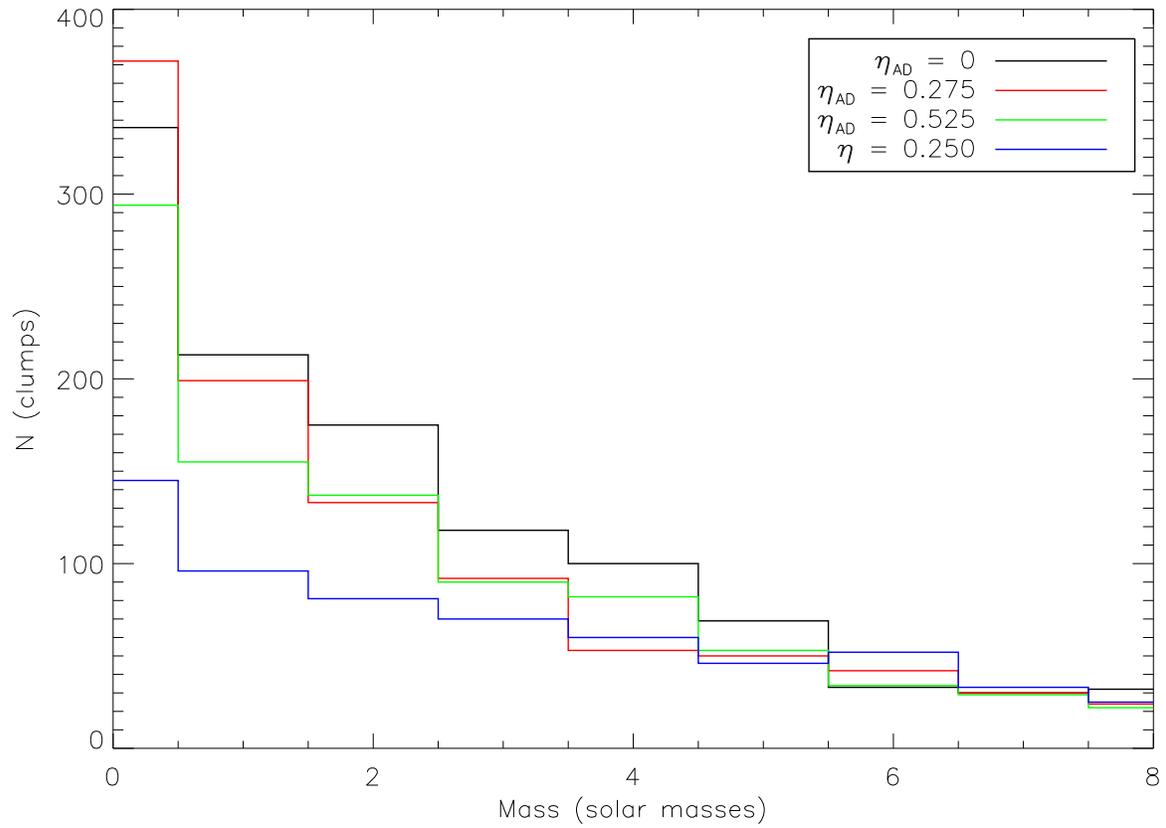}
\caption{Clump mass spectrum measuring the number of clumps of a given
  mass for one ideal MHD run (labeled $\eta_{AD} = 0$), two AD runs
  ($\eta_{AD} = 0.275, 0.525$) and one Ohmic dissipation run ($\eta =
  0.250$). Compare the lack of effect of AD to the significant
  decrease in the number of low mass clumps for the Ohmic diffusion
  case.
\label{clump_mass} }
\end{figure}

Nonetheless, density enhancements in a turbulent flow likely provide
the initial conditions for star formation.  To clarify the effects of
different turbulent dissipation mechanisms on the clump mass spectrum,
we study our three dimensional simulations of turbulence without
gravity. By using the {\sc clumpfind} algorithm
\citep*{1994ApJ...428..693W} on the density field to identify
contiguous regions of enhanced density, we can construct a clump mass
spectrum (Fig.~\ref{clump_mass}). Although such methods are
parameter-sensitive when attempting to draw comparisons to observed
estimates for the clump-mass spectrum \citep{2002ApJ...570..734B}, we
are only interested in using the mass spectrum as a point of
comparison between runs with different dissipative properties.

For this section, we dimensionalize our density field following 
\citet{1999ApJ...524..169M}, with a length scale $L' = 0.5$~pc, 
and mean density scale $\rho_0' = 10^4 (2 m_H)
\mbox{ g cm}^{-3}$ in order to present results in physical units
relevant to star formation.

We search for clumps above a density threshold set at $5 \langle \rho
\rangle$ (where in the AD cases $\rho = \rho_i + \rho_n$) and bin the
results by mass to produce a clump-mass
spectrum. Figure~\ref{clump_mass} shows that while Ohmic diffusion has
a dramatic effect on the number of low-mass clumps, AD has nearly
none. Although there are small fluctuations around the hydrodynamic
spectrum, there is no systematic trend with increasing strength of AD.
This result suggests that AD does not control the minimum mass of
clumps formed in turbulent molecular clouds.

\subsection{Magnetic Energy and Density Spectra}
The lack of an effect on the clump mass spectrum can be better
understood by examining the distribution of magnetic field and
density. 

\begin{figure} 
\plotone{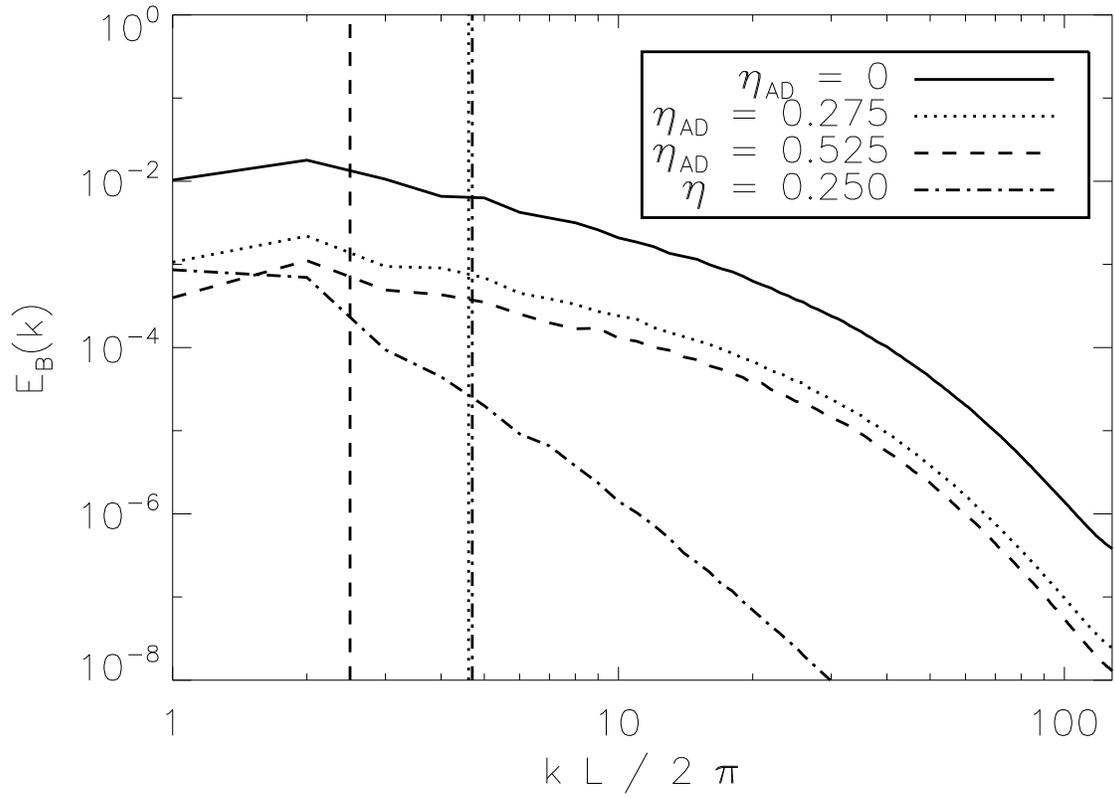}
\caption{Magnetic energy spectra for the same runs as
  Figure~\ref{clump_mass}. The vertical lines represent the wavenumber
  at which the AD or Ohmic Reynolds number crosses unity.
  \label{mag_spec}}
\end{figure}

AD produces no evident dissipation range in the magnetic energy
spectrum. As seen in Figure~\ref{mag_spec}, for two different values
of ambipolar diffusivity $\eta_{AD}$, the power spectrum of magnetic
field retains the shape of the ideal run.  For comparison, we have
also plotted the run with Ohmic diffusion. While the expected
dissipation wavenumbers (determined in both cases by the Reynolds
number method mentioned above) of the $\eta_{AD} = 0.275$ and $\eta =
0.250$ runs are very similar, the effect of Ohmic diffusion is quite
apparent in the declining slope of the magnetic energy spectrum, in
contrast to AD.

The total power does decrease as the ambipolar diffusivity $\eta_{AD}$
increases.  Because we drive only the neutrals, this could be
interpreted as magnetic energy being lost during the transfer of
driving energy from the ions to the neutrals via the
coupling. However, we performed a simulation in which both ions and
neutrals were driven with the same driving pattern and found almost no
difference in the power spectra from our standard (neutral driving
only) case.

We instead suspect that the decline in total magnetic energy occurs
because AD does damp some families of MHD waves, notably \al\ waves
\citep{1969ApJ...156..445K}, even though it does not introduce a
characteristic damping scale.

In order to demonstrate this, the flow will need to be decomposed into
its constituent MHD wave motions at each point in space. Such a
technique has been used before by \citet{2001ApJ...554.1175M} for
incompressible MHD turbulence and by \citet{2002PhRvL..88x5001C} for
compressible MHD turbulence. The technique used by
\citet{2002PhRvL..88x5001C} decomposes wave motions along a mean field
assumed to be present.  However, because the local field is distorted
by the turbulence and thus not necessarily parallel to the mean, a
mean-field decomposition tends to spuriously mix \al\ and slow modes
\citep{2001ApJ...554.1175M}. If the local field line distortion is
great enough, the decomposition must be made with respect to the local
field, a much more demanding proceedure. Although wave decomposition
analysis is outside the scope of this paper, it remains a fruitful
avenue for future research.

\begin{figure} 
\plotone{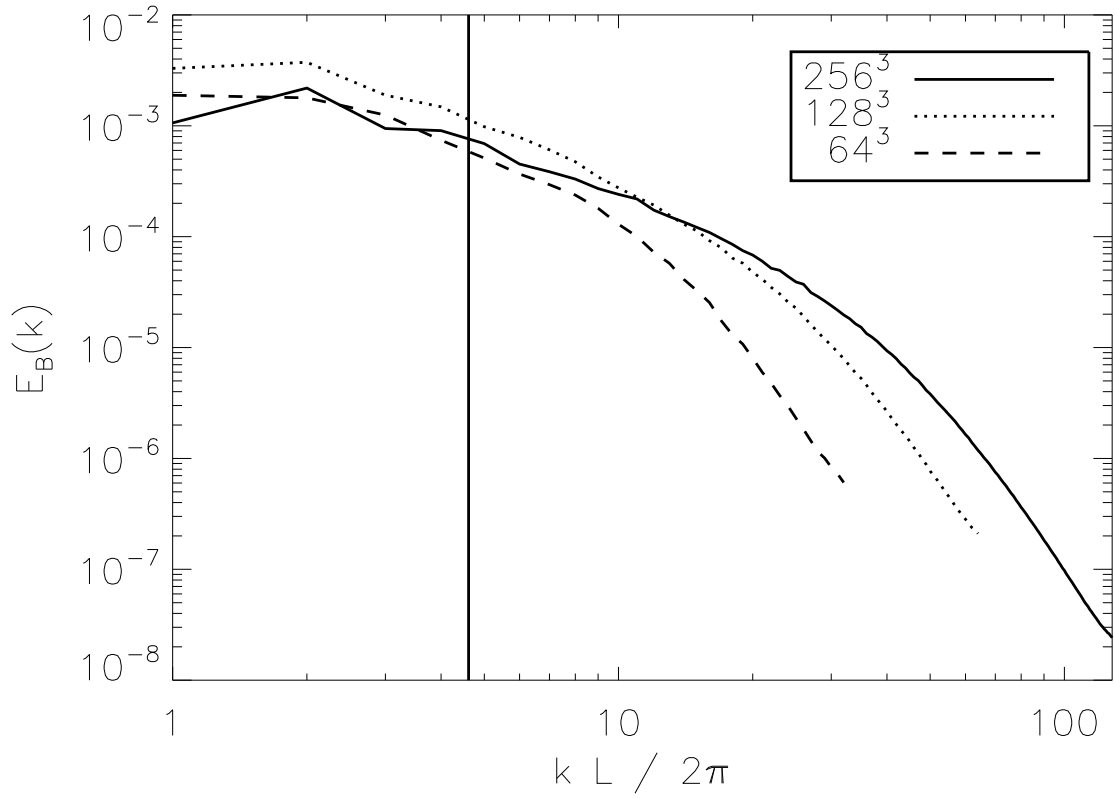}
\caption{Magnetic energy spectra for three runs of varying resolution
  from $64^3$ to $256^3$. Increased resolution shows no effects at the
  AD wavenumber given by the vertical line. 
\label{resolution} }
\end{figure}

In order to ensure that the lack of spectral features seen in the
magnetic spectrum (and similarly in the density spectrum) is not an
artifact of the limited inertial range in our simulations, we ran our
$\eta_{AD} = 0.275$ (medium collision strength) case at resolutions of
$64^3, 128^3,$ and $256^3$.  Figure~\ref{resolution} demonstrates that
increasing the resolution increases the inertial range, but does not
resolve any noticeable transition to dissipation at the AD length,
suggesting that our results are not sensitive to the resolution.

\begin{figure} 
\plotone{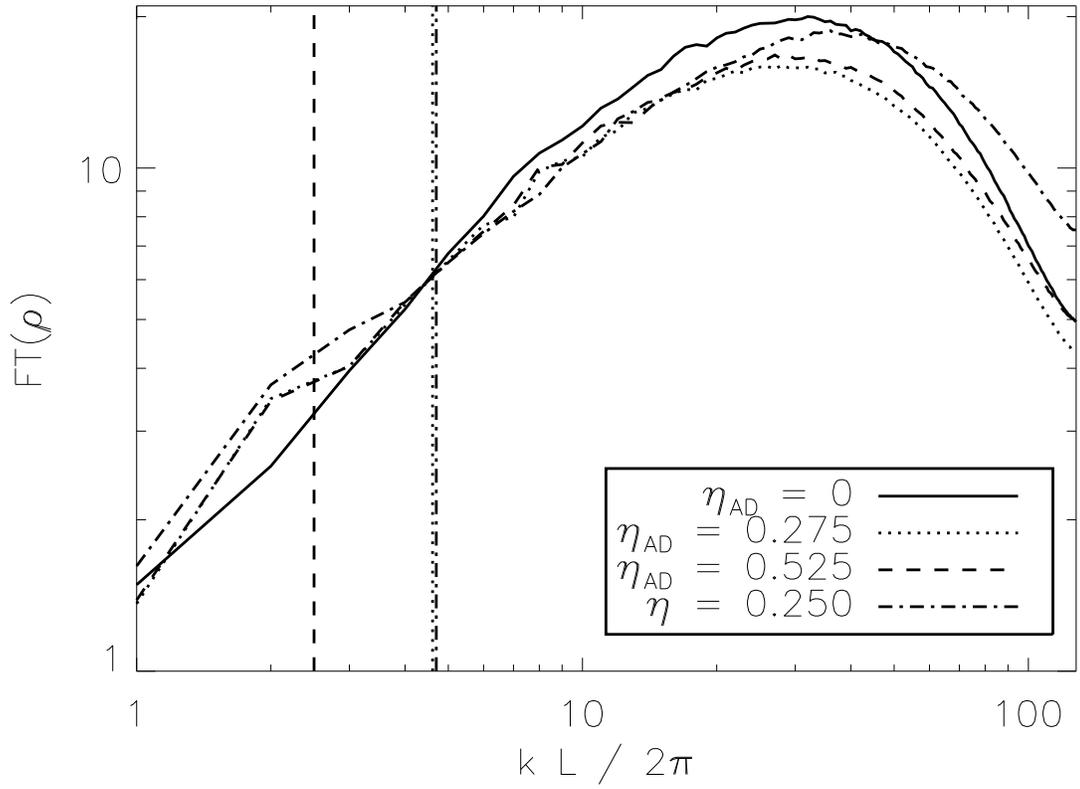}
\caption{Fourier transform of density field for the same runs as
    Figure~\ref{clump_mass}. 
  Note that for the AD runs, this plot shows the total density, $\rho
  = \rho_i + \rho_n$. 
  \label{rho_spec}}
\end{figure}

Figure~\ref{rho_spec} shows the spectrum of the density for all runs.
In the case of the AD runs, we use the sum of the neutral and ion
density.

The density spectrum peaks at small scale in compressible turbulence
\citep{jm05}.  Varying the ambipolar diffusivity by a factor of two
makes little systematic difference to the shape of the density
spectrum. It seems clear that although there are only slight
differences in the density spectrum due to varying magnetic
diffusivities, the density spectrum is not a particularly good
indicator of underlying clump masses.

Note that we use for the density spectrum the Fourier transform of the
density field rather than its square, which in the case of the
magnetic field yields the one-point correlation function (or power
spectrum) of the magnetic energy.

\section{Discussion}
Supersonic turbulence performs a dual role in its simultaneous ability
to globally support a molecular cloud against gravity while at the
same time producing smaller density enhancements that can sometimes
gravitationally collapse \citep{khm00}. While our simulations do not
include gravity, it is clear that AD does not set a characteristic
scale to the density field below which MHD turbulence is unable to
further influence structure formation.

One of the main motivations of this study was to verify the claim made
by, for example, \citet{khm00} that AD sets the minimum mass for
clumps in molecular cloud turbulence. However, it appears that AD is
unable to set this scale, because of its selective action on different
MHD waves. We do note that AD can occasionally help form
magnetohydrostatic objects in MHD turbulence, but this is not a
dominant pathway, as shown by \citet{2005ApJ...618..344V}. Although
Ohmic diffusion has little trouble inhibiting low mass clump
formation, it never reaches significant values at the densities where molecular
clumps form.

This opens up other possibilities for the physical mechanisms
determining the smallest scale fluctuations occurring in molecular
clouds. An attractive option is the sonic-scale argument of
\citet*{2003ApJ...585L.131V}, in which the length scale at which turbulent
motions become incompressible, with Mach numbers dropping well below
unity, determines where turbulence ceases to have an effect on the
pre-stellar core distribution, and thus determines the minimum mass scale.

\section{Acknowledgments}
We thank J. Ballesteros-Paredes and K. Walther for collaborating on
early phases of this work, and J. Maron, A. Schekochihin, J. Stone,
and E. Zweibel for productive discussions. We acknowledge support from
NASA grants NAG5-10103 and NAG5-13028. Computations were performed at
the Pittsburgh Supercomputer Center, supported by NSF, on an
Ultrasparc III cluster generously donated by Sun Microsystems, and on
the Parallel Computing Facility at the American Museum of Natural
History.

\clearpage

\clearpage

\clearpage

\clearpage

\clearpage

\clearpage

\begin{deluxetable}{rlllllll}
\tablecaption{
Models \label{run_tab}}
\tablehead{
\colhead{Run} & \colhead{diffusivity} & \colhead{$\gamma_{AD}$} &
\colhead{$\eta_{AD}$} & \colhead{$\eta$} & \colhead{$\sigma_{v,n}$} &
\colhead{$\sigma_{v,i}$} & \colhead{$\tau_{dr}$}
}
\startdata
A1 & AD & 8 & 0.275 & 0 & 0.603 & 0.526 & 1.66\\
A2 & AD & 4 & 0.575 & 0 & 0.615 & 0.501 & 1.63\\
O & Ohmic & 0 & 0 & 0.250 & - & 0.577 & 1.73\\
I & - & 0 & 0 & 0 & - & 0.630 & 1.59\\
\enddata
\end{deluxetable}

\end{document}